\documentclass[aps,prd,notitlepage,%
superscriptaddress,twocolumn]{revtex4-1}

\usepackage{natbib}
\usepackage{graphicx}
\usepackage{amssymb}
\usepackage{amsmath}
\usepackage{blkarray}
\usepackage{ifthen}
\usepackage{braket}
\usepackage{bm}
\usepackage{bbm}
\usepackage{comment}
\usepackage{siunitx}
\usepackage{float}
\usepackage{dcolumn}
\usepackage{cancel}
\usepackage{maybemath}

\newcolumntype{d}[1]{D{.}{.}{#1}}
\newcolumntype{.}{D{x}{}{9}}
\newcolumntype{,}{D{x}{}{5}}
\newcolumntype{;}{D{x}{}{19}}

\usepackage{fancyvrb}

\newcommand{\genspin}{S}

\newcommand{\dd}{\mathrm{d}}
\newcommand{\ii}{\mathrm{i}}
\newcommand{\ee}{\mathrm{e}}

\usepackage{bbm}

\begin{document}

\title{Tensor Polarizability of the Nucleus and Angular Mixing in
Muonic Deuterium}

\author{Gregory S. Adkins}
\affiliation{Franklin \& Marshall College,
Lancaster, Pennsylvania 17604, USA}

\author{Ulrich D. Jentschura}
\affiliation{Department of Physics and LAMOR, 
Missouri University of Science and
Technology, Rolla, Missouri 65409, USA}

\begin{abstract}
We investigate the effects of the tensor polarizability of a nucleus on the
bound-state energy levels, and obtain a general formula for the contribution of
the tensor polarizability to the energy levels in two-body bound systems.  In
particular, it is demonstrated that the tensor polarizability leads to mixing
between states with different orbital angular momenta.  The effect of tensor
polarizability is evaluated for the hyperfine-structure components of $P$
states and for the mixing of $S$ and $D$ states in muonic deuterium. 
\end{abstract}

\maketitle

%
% Introduction
%
\section{Introduction}
\label{sec1}

Nuclear effects are very important for muonic bound systems
in view of the smaller Bohr radius as compared to 
electronic bound systems~\cite{Pa2011,PaWi2015,%
KaPaYe2018,PaEtAl2024,JiZhPl2024}.
In an early investigation on the subject (Ref.~\cite{PaLeHa1993}),
it was shown that, for a deuteron nucleus, the 
displacement of the constituent proton (inside the 
deuteron) relative to the constituent neutron gives 
rise to an energy shift which, for $S$ states and 
in leading order, is proportional to the probability density 
at the origin. For non-$S$ states, 
because of the vanishing probability density at the 
origin, the effect enters in higher order.
In particular, for the $2P$ state of muonic deuterium,
the effect has been shown to be 
proportional to the product of the  nuclear polarizability constant 
and the matrix element
$\langle 1/r^4 \rangle$ of the bound muonic state
(here, $r$ is the muon-deuteron distance,
see Eqs.~(7) and~(10)
of Ref.~\cite{Pa2011}). The nucleus,
in this case, acts as a ``polarizable core''
which interacts with the
bound muon in much the same way
as a charged atomic core would otherwise
interact with a ``Rydberg electron''
(the latter point of view is illustrated 
in Chap.~6 of Ref.~\cite{JeAd2022book}).

In this work,  we explore an effect which 
transcends the simple picture of an electrically
polarizable deuteron and takes into 
account its spin structure, 
namely, the tensor polarizability.
In general, nuclei with spin
quantum numbers higher than spin-$1/2$ 
can exhibit a nonvanishing tensor polarizability.
Our aim is to derive a general formula
for the energy shift experienced by 
a bound muon (or electron) 
bound to a nucleus which has a nonvanishing 
tensor polarizability. One notes that, in contrast
to the scalar polarizability, the diagonal energy 
shift of $S$ states 
due to the nuclear tensor polarizability
vanishes after angular integration.
For non-$S$ states, we find that the effect of nuclear
tensor polarizability in muonic deuterium is small, not
yet discernible in experimental measurements,
but that it gives rise to a very interesting alteration
of the hyperfine mixing manifold, 
even leading (somewhat surprisingly) to mixed states
involving more than one value of the {\em orbital} angular
momentum.  For example, the nuclear tensor polarizability leads to a 
mixing of $S$ and $D$ states.

We use natural units with 
$\hbar = c = \epsilon_0 = 1$ throughout 
this paper, except where factors of $\hbar$, $c$, and $\epsilon_0$ 
are shown explicitly for emphasis.

%
% Polarizability Effect
%
\section{Tensor Polarizability}
\label{sec2}

We work with a generalized Bohr radius 
\begin{equation}
a_0 = \frac{\hbar}{Z \alpha m_r c} \,,
\end{equation}
where $Z$ is the nuclear charge number,
$\alpha$ is the fine-structure constant and
$m_r$ is the reduced mass.
With Friar and Payne~\cite{FrPa2005deuteron},
we assume that the polarizability tensor
$(\alpha_P)^{i j}$ of the nucleus 
can be expressed as 
\begin{equation}
\label{pol012}
(\alpha_P)^{i j} = \alpha_E \frac{\delta^{i j}}{3} +
\ii \, \sigma_N \, \epsilon^{i j k}
\frac{\genspin_N^k}{2} + \tau_N ( \genspin_N^i \genspin_N^j )^{(2)} \,.
\end{equation}
Here, 
the constants $\alpha_E$, $\sigma_N$ and $\tau_N$ parameterize the
scalar ($\ell = 0$), vector ($\ell = 1$) and tensor ($\ell = 2$) 
components of the nuclear polarizability, respectively.
The second-rank nuclear spin tensor is 
\begin{equation}
S^{ij}_{(2)} = ( S_N^i S_N^j )_{(2)} =
\tfrac12 ( S_N^i S_N^j + S_N^j S_N^i ) -
\tfrac{\delta^{ij}}{3} \, (\vec S_N)^2 \,.
\end{equation}
We denote Cartesian indices by superscripts 
and use the Einstein summation convention.
For the (static) scalar and tensor polarizabilities of the deuteron,
the calculated results of Friar and Payne~\cite{FrPa2005deuteron}
read as follows,
\begin{subequations}
\label{alphaEtaud}
\begin{eqnarray}
\label{alphaE}
\alpha_E &=& 0.6330(13) \, {\rm fm}^3 \, , \\
\label{taud}
\tau_{N=d} &=& 0.0317(3) \, {\rm fm}^3 \,.
\end{eqnarray}
\end{subequations}
These results are expressed in terms of a polarization
volume, {\em i.e.}, in units of fermi cubed.
For reference, we indicate the conversion of the 
polarization volume to the SI polarizability,
which, for the scalar polarizability, reads as follows,
\begin{equation}
\label{alphaSI}
( \alpha_E)_{\rm SI} \,
= 4 \pi \epsilon_0 \, (\alpha_E)_{\rm vol} \, ,
\end{equation}
where $\epsilon_0$ is the vacuum permittivity.
The numerical values in Eq.~\eqref{alphaEtaud}
have to be interpreted in terms of volume 
polarizabilities.

A comprehensive derivation of the polarizability
correction of an atomic core for non-$S$ states
of a bound system is presented
in Sec.~6.6 of Ref.~\cite{JeAd2022book}.
A generalization of
Eq.~(6.174) of Ref.~\cite{JeAd2022book} for a polarizable
nucleus shows that, for non-$S$ states,
\begin{multline} 
\label{general_polarizability}
E_{\rm pol} = - \frac{3 \alpha}{2} 
\left< \frac{\hat x^i}{r^2}
\, \left( \alpha_E \frac{\delta^{ij}}{3} \, 
+ \tau_N \, ( S_N^i \, S_N^j )^{(2)}
\right) \, \frac{\hat x^j}{r^2} \right>
\\
= - \frac{\alpha_E}{2} \left< \frac{\alpha}{r^4} \right> 
- \frac{3 \tau_N }{2} 
\left< X^{ij}_{(2)} \, S^{ij}_{(2)} \frac{\alpha}{r^4} \right> \,,
\end{multline}
where we use the Einstein summation
convention, and the second-rank coordinate tensor is
\begin{equation}
\label{tensordef}
X^{ij}_{(2)} = (\hat x^i \, \hat x^j)_{(2)} =
\hat x^i \, \hat x^j  - \tfrac13 \delta^{ij} \,,
\end{equation}
The vector-polarizability term proportional to $\sigma_N \, \epsilon^{i j k}$
in Eq.~\eqref{pol012} does not contribute to
$E_{\rm pol}$ because $\hat x^i \, \hat x^j \,  \epsilon^{i j k} = 0$.
The coupling of angular momenta in
one-muon ions proceeds by first 
coupling the orbital angular momentum $\vec L$
and the spin $\vec S_\mu$ of the muon
to form the total
angular momentum of the muon
according to $\vec J = \vec L + \vec S_\mu$.
One then couples $\vec J$ to the 
nuclear spin $\vec S_N$, according to 
$\vec F = \vec J + \vec S_N$.
This is appropriate because the spin-orbit
coupling of the bound muon is stronger than the
spin--spin coupling of the muon
and the nucleus.

In a one-muon ion, one
parameterizes the bound states as follows
(with hyperfine resolution)
\begin{subequations}
\begin{equation}
\psi^{F M_F}_{n L S_\mu J S_N}(\vec r \, ) = R_{nL}(r) \,
\Xi^{F M_F}_{L S_\mu J S_N}(\theta, \varphi) \,, 
\end{equation}
where the hyperfine-resolved
spin-angular function is
\begin{multline}
\Xi^{F M_F}_{L S_\mu J S_N}(\theta, \varphi) =
\sum_{m_\mu m_N}
C^{F M_F}_{S_N m_N J M_J} \;
C^{J M_J}_{L m_L S_\mu m_\mu} 
\\
\times
Y_{L m_L}(\theta, \varphi) \;
\chi^{(\mu)}_{m_\mu} \;
\chi^{(N)}_{m_N} \,,
\end{multline}
\end{subequations}
where $M_J=M_F-m_N$ and $m_L=M_J-m_\mu$.
The fundamental spinors for the 
muon [superscript $(\mu)$] and the 
nucleus [superscript $(N)$] are
denoted as $\chi^{(\mu)}_{m_\mu}$ and
$\chi^{(N)}_{m_N}$, respectively.
Via angular reduction formulas, which are investigated in the next 
section, one arrives at the following, general result
for the matrix element of an operator
proportional to $f(r) \, X_{(2)}^{ij} \, S_{(2)}^{ij}$,
taken with hyperfine-resolved bound-state wave functions,
\begin{multline}
\int \dd^3 r \,
[\psi^{F M_F}_{n L' S_\mu J' S_N}(\vec r\, )]^\dagger \,
f(r) \, X_{(2)}^{ij} \, S_{(2)}^{ij}
\psi^{F M_F}_{n L S_\mu J S_N}(\vec r \, ) \\
= G^{S_\mu S_N F}_{L' J'; L J} \int \dd r \, r^2 \,
R_{nL'}(r)  \, f(r) \, R_{n L}(r) \,.
\end{multline}
Here, $G^{S_\mu S_N F}_{L' J'; L J}$ is 
an angular factor calculated in Eq.~\eqref{input1}.
We take into account the mixing 
of the fine-structure and hyperfine structure
by allowing for different orbital angular 
momentum numbers $L'$, $L$ and total muon angular 
momentum numbers $J'$, $J$ in the 
bra and ket states. Finally, in the 
manifold of states with given $F$ and $M_F$,
the matrix of the energy perturbations by the nuclear
polarizability is 
\begin{subequations}
\label{EEmatrix}
\begin{align}
( E_{\rm pol} )_{L' J'; L J}  = & \;
( E_\alpha )_{L' J'; L J}  +
(E_\tau )_{L' J'; L J}  \,,
\\[2ex]
( E_\alpha )_{L' J'; L J}  = & \;
-\frac12 \alpha_E \, 
\delta_{L' L} \, \delta_{J'J}
\left< \frac{\alpha}{r^4} \right>_{L L}  \,,
\\[2ex]
( E_\tau )_{L' J'; L J} = & \;
-\frac{3}{2} \, \tau_N \,
G^{S_\mu S_N F}_{L' J'; L J}  \,
\left< \frac{\alpha}{r^4} \right>_{L' L} \, .
\end{align}
\end{subequations}
%
%A few general remarks on the angular
%factor $G^{S_\mu S_N F}_{L' J'; L J}$ are in order.
We note that parity conservation (in the absence of weak interactions) implies that the
angular factor $G^{S_\mu S_N F}_{L' J'; L J}$ vanishes for values of $L'$ and $L$ that differ by an odd
integer, as can be confirmed from the explicit expression in \eqref{input1}. 

In regard to the leading hyperfine interactions (spin-orbit, spin-spin, and
quadrupole terms), one notices that the quadrupole terms 
could in principle couple states where $L'$ and $L$ differ by an even integer;
however, the corresponding radial matrix elements
are proportional to $1/r^3$ and vanish when $L'$ and $L$ differ by an even integer, 
as noticed by Pirenne \cite{Pi1947iii}.  So, the complete matrix elements of these
hyperfine interactions are diagonal in the orbital angular momentum quantum number.  By
contrast, the matrix elements of $1/r^4$ are non-zero for values of $L'$ and
$L$ that differ by an even integer and must be taken into account in the
expression for the tensor polarizability matrix elements.  The radial matrix
element is given explicitly in terms of an integral over associated Laguerre
polynomials as 
\begin{multline} 
\label{onebyr4_integral} 
\left \langle
\frac{1}{r^4} \right \rangle_{L' L} = \frac{8 (m_r Z \alpha)^4}{n^5} 
\left ( \frac{p'! \, p!}{q'! \, q!} \right )^{1/2} \\ 
\times \int_0^\infty d \rho \,
\rho^{L'+L-2} \ee^{-\rho} L_{p'}^{s'}(\rho) L_p^s(\rho) \, , 
\end{multline} 
where $p=n-L-1$, $q=n+L$, and $s=2L+1$, and
$p'=n-L'-1$, $q'=n+L'$, and $s'=2L'+1$, {\em i.e.},
$p'$, $q'$, and $s'$ are obtained from the 
corresponding $p$, $q$, and $s$ by replacing $L$ with $L'$.
The integral in Eq.~\eqref{onebyr4_integral} 
is of the Gordon type~\cite{Go1929aop},
and can be expressed in terms of the 
Appell $F_2$ function (see Eqs.~(1), (6) and (17) 
of Ref.~\cite{Ma1991appell}).
Alternative representations in terms of 
the Appell $F_1$ and $F_3$ functions have been
discussed in Ref.~\cite{Ta2002}. Here, for definiteness, we leave the 
integral in the form given in Eq.~\eqref{onebyr4_integral},
because its evaluation, for given quantum numbers,
is straightforward using modern computer algebra systems~\cite{Wo2024}.

%
% Tensor Structures}
%
\section{Tensor Algebra}
\label{sec3}

In this section, we derive an expression for the matrix elements $G^{S_1 S_2
F}_{L' J'; L J} $ of the polarizability operator $X^{i j}_{(2)} S^{i j}_{(2)}$
where the second rank tensors $X^{i j}_{(2)}$ and $S^{i j}_{(2)} = ( S_2^i
S_2^j )_{(2)}$ are defined in \eqref{tensordef}.  We give a general derivation
appropriate for systems of any spin values.

The angular momenta relevant for two-body bound systems are the spins $\vec
S_1$ and $\vec S_2$ of the two particles and the orbital angular momentum $\vec
L$.  For systems like muonic deuterium having two bodies with significantly
different masses $m_1$ and $m_2$ (with $m_1 < m_2$), the appropriate angular
momentum coupling scheme combines first the orbital $\vec L$ with the spin of
the less massive particle $S_1$ to form a subtotal angular momentum $\vec J =
\vec L + \vec S_1$, which is then combined with the spin of the more massive
particle to form the total angular momentum $\vec F = \vec J + \vec S_2$.  This
coupling scheme could be represented as $L S_1 J S_2 F$, or briefly as the $L J
F$ scheme.  Angular states in this scheme can be constructed from
Clebsch-Gordan coefficients as
\begin{multline}
\vert L S_1 J S_2 F M \rangle = \sum_{S_{1 z} S_{2z}} C^{F M}_{J J_z; S_2 S_{2z}} \, C^{J J_z}_{L L_z; S_1 S_{1z}} \, \\
\times \vert L L_z \rangle \, \vert S_1 S_{1z} \rangle \, \vert S_2 S_{2z} \rangle
\end{multline}
where $M\equiv F_z$ is the $z$-component of the total angular momentum, $J_z
\equiv M-S_{2z}$, and $L_z \equiv M-S_{1z}-S_{2z}$.  The orbital states could
be represented as spherical harmonics $\vert L L_z \rangle \rightarrow Y_{L
L_z}(\theta, \phi)$ and the spin states as two-component Pauli spinors for
spin-1/2, etc.

The spherical tensors $\vec X_{(2)}$ and $\vec S_{(2)}$ corresponding to $X^{i
j}_{(2)}$ and $S^{i j}_{(2)}$ are normalized so that
\begin{align}
X^{(2)}_{20} =& \;
\sqrt{\frac32} (\hat x^i \hat x^j)_{20} =
\sqrt{\frac32} \, 
(\cos^2 \theta - \tfrac13) \,,
\\
S^{(2)}_{2 0} =& \;
\sqrt{\frac32} \,
\big [ S_2^{i=3} \, S_2^{j=3} - \frac13 (\vec S_2)^2 \big ] \,.
\end{align}
where we denote Cartesian indices as superscripts and
spherical tensor indices as subscripts.
This normalization of the spherical tensors follows from the use
of the Clebsch--Gordan coefficients, as explained
in Chap.~6 of Ref.~\cite{JeAd2022book}.
One can show the following relation between the 
Cartesian and spherical formulations,
\begin{equation}
X_{(2)}^{ij} \, S_{(2)}^{ij} = \vec X_{(2)} \cdot \vec S_{(2)} \equiv \sum_{q=-2}^2
(-1)^q \, X^{(2)}_{2,q} \, S^{(2)}_{2,-q}  \,.
\end{equation}
The $\vec X_{(2)}$ tensor acts on the orbital angular momentum $\vec L$, 
while the tensor $\vec S_{(2)}$ acts on the spin angular momentum $\vec S_2$
of particle 2 (the nucleus).
Our goal is to calculate
\begin{equation}
G^{S_1 S_2 F}_{L' J'; L J} 
\equiv \langle L' S_1 J' S_2 F M | \, X_{(2)}^{ij} \, 
S_{(2)}^{ij} | L S_1 J S_2 F M  \rangle  \,.
\end{equation}
The expression $G^{S_1 S_2 F}_{L' J'; L J} $
is independent of the magnetic projection $M$.
Angular reduction according to Eq.~(7.1.6) of Ref.~\cite{Ed1957}
leads to 
%
% j'_1 = J'
% j'_2 = S_N
% j_1 = J
% j_2 = S_N
% J' = J = F
%
\begin{multline} 
\label{input3}
\langle L' S_1 J' S_2 F' M' | \, X_{(2)}^{ij} \, 
S_{(2)}^{ij} | L S_1 J S_2 F M  \rangle \\
= \delta_{F' F} \delta_{M' M} (-1)^{ J + S_2 + F} \, 
\left\{ \begin{array}{ccc} F & S_2 & J' \\ 2 & J & S_2 \end{array} \right\} 
\\
\times
\langle L' S_1 J' || \vec X_{(2)} || L S_1 J \rangle \,
\, \langle S_2 || \vec S_{(2)} || S_2 \rangle \,.
\end{multline} 
According to Eq.~(7.1.7) of Ref.~\cite{Ed1957},
one can further reduce the first reduced matrix element to
\begin{multline}
\langle L' S_1 J' || \vec X_{(2)} || L S_1 J \rangle =
(-1)^{L'+ S_1 + J} \, \Pi_{J' J} \\
\times
\left\{ \begin{array}{ccc} L' & J' & S_1 \\
J & L & 2 \end{array} \right\} \,
\langle L' || \vec X_{(2)} || L \rangle \, ,
\end{multline}
where $\Pi_{a b} \equiv \sqrt{(2a+1) \; (2b+1)}$.  
Finally, one has the result
\begin{multline}
\label{input2}
G^{S_1 S_2 F}_{L' J'; L J} 
= (-1)^{2J+L' + S_1+S_2+F } \, \Pi_{J' J} \\
\times \left\{ \begin{array}{ccc} F & S_2 & J' \\ 2 & J & S_2 \end{array} \right\} 
\left\{ \begin{array}{ccc} L' & J' & S_1 \\ J & L & 2 \end{array} \right\}  \\
\times \langle L' || \vec X_{(2)} || L \rangle \, 
\langle S_2 || \vec S_{(2)} || S_2 \rangle \,.
\end{multline}
The reduced matrix element for the angular tensor is 
[see Eq.~(6.69) of Ref.~\cite{JeAd2022book}] 
\begin{equation} 
\label{input4}
\langle L' || \vec X_{(2)} || L \rangle = 
\sqrt{\frac{2}{3}} (-1)^{L'} \Pi_{L' L} 
\left( \begin{array}{ccc} L' & 2 & L \\
0 & 0 & 0 \end{array} \right) \,.
\end{equation} 
For the second-rank nuclear spin-tensor, we have
\begin{equation} 
\langle S_2 || \vec S_{(2)} || S_2 \rangle 
= \sqrt{\frac{1}{24}} \sqrt{\frac{(2S_2+3)!}{(2S_2-2)!}} \, \Theta(S_2 \ge 1)
\end{equation} 
as a straightforward consequence of the Wigner-Eckhart theorem 
(see (6.39) of Ref.~\cite{JeAd2022book}). For absolute
clarity and reference, we mention that the Heaviside $\Theta$ vanishes for $S_2 = 1/2$, 
in view of the fact that it is impossible to construct a spin-$2$ 
tensor from just two spin-$1/2$ objects,
while the Heaviside function is equal to unity for  
other integer as well as half-integer values
greater than $1/2$ ({\em i.e.}, $S_2 = 1, \tfrac32, 2, \dots$).
In summary, one obtains 
\begin{multline}
\label{input1}
G^{S_1 S_2 F}_{L' J'; L J} 
= (-1)^{2J+S_1+S_2+F } \, \Pi_{J' J} \Pi_{L' L}
\left\{ \begin{array}{ccc} F & S_2 & J' \\ 2 & J & S_2 \end{array} \right\} \, \\
\times \left\{ \begin{array}{ccc} L' & J' & S_1 \\ J & L & 2 \end{array} \right\} 
\left( \begin{array}{ccc} L' & 2 & L \\ 0 & 0 & 0 \end{array} \right) \,
\frac{1}{6} \sqrt {\frac{(2S_2+3)!} {(2S_2-2)!}} \Theta(S_2 \ge 1) \, .
\end{multline}

%
% Muonic Deuterium
%
\section{Polarizability and Mixing in Muonic Deuterium}
\label{sec4}

%
% Orientation
%
\subsection{Scalar and Tensor Effects}

Let us consider the effect of polarizability in a few interesting cases,
taking muonic deuterium as an example.  The expression
\begin{equation}
E_\alpha = - \frac{\alpha_E}{2} \left \langle \frac{\alpha}{r^4} \right \rangle 
\end{equation}
from \eqref{general_polarizability} for the scalar polarizability correction
gives a divergent result for $S$ states, and indeed the effect of polarizability
in S states requires a more careful discussion of the deuteron internal
structure 
\cite{FrPa2005deuteron,Pa2011,PaWi2015,KaPaYe2018,PaEtAl2024}. 
The approximation $E_\alpha$ for the
scalar polarizability gives a finite contribution for states with $L>0$,
\begin{multline}
E_\alpha = -2 \tilde \alpha_E \frac{m_r \alpha^5}{n^5} 
\left( \frac{m_r}{m_d} \right )^3 \\ 
\times \frac{3n^2-L(L+1)}{L (L+1) (2L-1) (2L+1) (2L+3)} \,,
\end{multline}
where $m_r$ is the muonic deuterium reduced mass and the 
dimensionless scalar polarizability constant is
\begin{equation} 
\label{alphaEdimless}
\tilde \alpha_E = \frac{m_d^3 \, \alpha_E}{(\hbar c)^3} = 543.6(1.1) \, .
\end{equation}
The scalar polarizability gives a contribution to the 2P energy level of 
\begin{equation}
\label{c1term}
E_\alpha (2P) = - \frac{\alpha_E}{2} 
\frac{ \alpha (m_r \alpha)^4}{24} = -3.555(8) \, \mu{\rm eV} \, .
\end{equation}
This contribution can be identified as the leading
term in the  Coulomb corrections to the
electric dipole polarizability 
when considering the expansion of the muonic dipole matrix
element given in Eq.~(9) of Ref.~\cite{PaWi2015}. 
In fact, the term in Eq.~\eqref{c1term} corresponds to the $c_1$
term from Ref.~\cite{PaWi2015}, where $c_1(2P) = 1/24$ as 
given in Eq.~(7) of Ref.~\cite{Pa2011}.  
The expression for $\alpha_E$ as an energy integral is given in
Eq.~(10) of Ref. \cite{Pa2011}.  Contributions of order $4 \, \mu{\rm eV}$ are
roughly the size of current experimental uncertainties for the 
$2S$--$2P$ intervals
in muonic deuterium~\cite{PoEtAl2016}.  Prospects for the observation of scalar
polarizability in heavy hadronic atoms has recently been discussed in
Ref.~\cite{LiOhShSo2025}.

As discussed above, the tensor polarizability alters the mixing scheme for
hyperfine states, leading to mixing of states having different values of
orbital angular momentum $L$ as well as different values of the total muon
angular momentum $J$. In terms of magnitude, the contribution 
of the tensor polarizability is not parametrically suppressed
by higher powers of the fine-structure constant
in comparison to the scalar polarizability. 
However, when considering Eq.~\eqref{alphaEtaud} for the deuteron 
as a guide for other nuclei, one could conjecture
that, in general, typical tensor contributions might be expected
to be roughly 20 times smaller than scalar ones due to the sizes of the
coefficients encountered for 
$\alpha_E$ and $\tau_N$. For context, one should remember, though,
that the numerical value for $\tau_d$ given in Eq.~\eqref{taud} is based on 
data-driven calculations and not direct measurements.

%
% 2P
%
\subsection{Mixing of $\maybebm{2P}$ States of Muonic Deuterium}
\label{sec4a}

The overall structure of mixing of $2P$ states is not affected by the tensor
polarizability, which only makes small corrections to the matrix elements.  The
mixed states have defined values of total angular momentum $F$ but possibly
different values of the total muon angular momentum $J$ (where $\vec J = \vec L
+ \vec S_\mu$).  For $n=2$ and $L=1$, the mixed states are $\{
2P_{1/2}^{F=1/2}, \, 2P_{3/2}^{F=1/2} \}$ and $\{ 2P_{1/2}^{F=3/2}, \,
2P_{3/2}^{F=3/2} \}$, while $2P_{3/2}^{F=5/2}$ is unmixed.  The matrix of
angular factors $G^{S_\mu S_d F}_{L J'; L J} $ in the $2P$-state manifold is
[we use the same ordering of states as
in Eq.~(45) of Ref.~\cite{KrEtAl2016deuterium}]
\begin{widetext}
\begin{equation}
\label{GGmatrix}
\mathbbm{G} =
\bordermatrix{
~ & 2P_{1/2}^{F=1/2} & 2P_{1/2}^{F=3/2} &
2P_{3/2}^{F=1/2} & 2P_{3/2}^{F=3/2} & 2P_{3/2}^{F=5/2} \cr
2P_{1/2}^{F=1/2} & 0 & 0 & -\frac{\sqrt{2}}{3} & 0 & 0 \cr
2P_{1/2}^{F=3/2} & 0 & 0 & 0 & \frac{1}{3 \sqrt{5}} & 0 \cr
2P_{3/2}^{F=1/2} & -\frac{\sqrt{2}}{3} & 0 & -\frac13 & 0 & 0 \cr
2P_{3/2}^{F=3/2} & 0 & \frac{1}{3 \sqrt{5}} & 0 & \frac{4}{15} & 0 \cr
2P_{3/2}^{F=5/2} & 0 & 0 & 0 & 0 & -\frac{1}{15} \cr } \,.
\end{equation}

\end{widetext}
Taking into account the multiplicity
of states $2F+1$ with a given value of $F$, we note that the 
trivial algebraic identity 
$2 \times (-1/3) + 4 \times \frac{4}{15} + 6 \times (-1/15) = 0$ 
implies that the multiplicity-weighted (hyperfine-averaged)
trace of the $\mathbbm{G}$ matrix in Eq.~\eqref{GGmatrix}
(and, thus, the sum of its eigenvalues) vanishes.

In terms of adjacency, one observes that the 
coupling happens between states of different $J$, but the same $L$ and $F$.  
We can thus separate the matrices as follows,
\begin{equation}
\mathbbm{G}(F=\tfrac12) =
\bordermatrix{
~ & 2P_{1/2}^{F=1/2} & 2P_{3/2}^{F=1/2} \cr
2P_{1/2}^{F=1/2} & 0 & -\frac{\sqrt{2}}{3} \cr
2P_{3/2}^{F=1/2} & -\frac{\sqrt{2}}{3} & -\frac13 \cr } \,,
\end{equation}
and
\begin{equation}
\mathbbm{G}(F=\tfrac32) = \bordermatrix{
~ & 2P_{1/2}^{F=3/2} & 2P_{3/2}^{F=3/2} \cr
2P_{1/2}^{F=3/2} & 0 & \frac{1}{3 \sqrt{5}} \cr
2P_{3/2}^{F=3/2} & \frac{1}{3 \sqrt{5}} & \frac{4}{15} 
\cr } \, . 
\end{equation}
The matrix for $F=5/2$ is a $1 \times 1$ matrix and is 
disconnected from the rest.
We note that the average eigenvalue for the $F=1/2$ states
(the trace of the $2 \times 2$ matrix) is $-1/3$, and the average for the
$F=3/2$ states is $4/15$.

Tensor polarizability contributions to the hyperfine energies are
given by

\begin{eqnarray}
(E_\tau)_{L' J';L J} &=& -\frac{3}{2} \tau_N G^{S_\mu S_d F}_{L' J'; L J} 
\left \langle \frac{\alpha}{r^4} \right \rangle_{L' L} 
\nonumber \\
&\phantom{x}& \hspace{-1.4cm} = 
- 12 \; \tilde \tau_N \;
\frac{m_r \alpha^5}{n^5} \left ( \frac{m_r}{m_d} \right )^3
G^{S_\mu S_d F}_{L' J'; L J} \, I^n_{L' L} \, ,
\end{eqnarray}
where
\begin{equation} \label{tauN}
\tilde \tau_N = \frac{m_d^3 \, \tau_N}{(\hbar c)^3} = 27.22(26) 
\end{equation}
is the dimensionless tensor polarizability constant, 
and the radial integral factor [see Eq.~\eqref{onebyr4_integral}
and the definitions in the text following Eq.~\eqref{onebyr4_integral}] is
\begin{equation}
I^n_{L' L}  \equiv \left ( \frac{p'! \, p!}{q'! \, q!} \right )^{1/2} 
\int_0^\infty \dd \rho \, \rho^{L'+L-2} \ee^{-\rho} 
L_{p'}^{s'}(\rho) L_p^s(\rho) \, .
\end{equation}

The matrices $\mathbbm{E}_{\tau}$ 
of energy shifts read as follows,
\begin{equation}
\frac{\mathbbm{E}_{\tau}(F=\tfrac12)}{\mu\mathrm{eV} }  =
\bordermatrix{
~ & 2P_{1/2}^{F=1/2} & 2P_{3/2}^{F=1/2} \cr
2P_{1/2}^{F=1/2} & 0 & 0.252(3) \cr
2P_{3/2}^{F=1/2} & 0.252(3) & 0.178(2) \cr } \, \,,
\end{equation}
and
\begin{equation}
\frac{\mathbbm{E}_{\tau}(F=\tfrac32)}{\mu\mathrm{eV} }  =
\bordermatrix{
~ & 2P_{1/2}^{F=3/2} & P_{3/2}^{F=3/2} \cr
2P_{1/2}^{F=3/2} & 0 & -0.0796(8) \cr
2P_{3/2}^{F=3/2} & -0.0796(8)  & -0.142(2) \cr } \,  \,.
\end{equation}
The correction for the $2P_{3/2}^{F=5/2}$ state is
\begin{equation}
E_\tau (F= \tfrac52) = 0.0356(4) \, \mu{\rm eV} \, .
\end{equation}

%
% 3S/3D
%
\subsection{Mixing of $\maybebm{3S}$ and $\maybebm{3D}$ States}
\label{sec4b}

The new feature that tensor polarizability introduces into the theory of
the hyperfine structure is the inclusion of states with different values
of orbital angular momentum $L$ into the same hyperfine multiplet.  
This effect first manifests for $n=3$ where there is mixing
of S and D states.  The matrices of angular factors
$G^{S_\mu S_d F}_{L' J'; L J} $ in the coupled
$(3S;3D)$-state manifold are (for $F=1/2$)
\begin{equation}
\mathbbm{G}(F = \tfrac12) =
\bordermatrix{
~ & 3S_{1/2}^{F=1/2} & 3D_{3/2}^{F=1/2} \cr
3S_{1/2}^{F=1/2} & 0 & -\tfrac{\sqrt{2}}{3} \cr
3D_{3/2}^{F=1/2} & -\tfrac{\sqrt{2}}{3} & -\tfrac13 \cr } \,,
\end{equation}
while for $F = 3/2$, the matrix is $3 \times 3$ and reads as follows
(in the basis spanned by the states
$3S_{1/2}^{F=3/2}$, $3D_{3/2}^{F=3/2}$ and $3D_{5/2}^{F=3/2}$)
\begin{equation}
\mathbbm{G}(F = \tfrac32) =
\left( \begin{array}{ccc} 
0 & \tfrac{1}{3 \sqrt{5}} & \frac{1}{\sqrt{5}} \\[2ex]
\tfrac{1}{3 \sqrt{5}} & \tfrac{4}{15} & -\tfrac{1}{5} \\[2ex]
\tfrac{1}{\sqrt{5}} & -\tfrac15 & -\frac{4}{15} 
\end{array} \right) \,.
\end{equation}
Finally, for $F = 5/2$, one obtains
\begin{equation}
\mathbbm{G}(F = \tfrac52) =
\bordermatrix{
~ & 3D_{3/2}^{F=5/2} & 3D_{5/2}^{F=5/2} \cr
3D_{3/2}^{F=5/2} & -\frac{1}{15} & \tfrac{\sqrt{2}}{5\sqrt{7}} \cr
3D_{5/2}^{F=5/2} & \tfrac{\sqrt{2}}{5 \sqrt{7}} & \tfrac{32}{105} \cr } \, .
\end{equation}
The angular factor for the $3D_{5/2}^{F=7/2}$ state is $-\frac{2}{21}$.
In view of the trivial algebraic identity,
\begin{equation}
2 \left ( -\tfrac13 \right ) + 
4 \left ( \tfrac{4}{15}-\tfrac{4}{15} \right ) +
6 \left ( - \tfrac{1}{15} + \tfrac{32}{105} \right ) - 
8 \left ( \tfrac{2}{21} \right ) = 0 \,,
\end{equation}
we confirm that the hyperfine-averaged energy shift 
for the $(3S;3D)$-state manifold vanishes. 
Off-diagonal $I^n_{L' L}$ factors with $L' \neq L$
do not affect the trace of the $\mathbbm{G}$ matrix
and, hence, do not change this conclusion.
Also, the radial matrix element does not
affect this conclusion, since all contributing states have
$n=3$ and $L=2$, so the radial matrix 
element is the same for all states contributing to the trace.
The energy matrices for the $(3S;3D)$-state 
manifold are as follows. For $F = 1/2$, one obtains
\begin{equation}
\frac{\mathbbm{E}_\tau(F = \tfrac12)}{\mu\mathrm{eV}} =
\bordermatrix{
~ & 3S_{1/2}^{F=1/2} & 3D_{3/2}^{F=1/2} \cr
3S_{1/2}^{F=1/2} & 0 & 0.0105(1) \cr
3D_{3/2}^{F=1/2} & 0.0105(1) & 0.00234(2) \cr } \, \,.
\end{equation}
For $F = 3/2$, the corresponding $3 \times 3$ matrix is
\begin{widetext}
\begin{equation}
\mathbbm{E}_\tau(F = \tfrac32) =
\bordermatrix{
~ & 3S_{1/2}^{F=3/2} & 3D_{3/2}^{F=3/2} & 3D_{5/2}^{F=3/2} \cr
3S_{1/2}^{F=3/2} & 0 & -0.00332(3) & -0.00995(10) \cr
3D_{3/2}^{F=3/2} & -0.00332(3) & -0.00188(2) & 0.00141(2) \cr
3D_{5/2}^{F=3/2} & -0.00995(10) & 0.00141(2) & 0.00188(2) \cr } \, 
  \mu\mathrm{eV} \, .
\end{equation}
\end{widetext}
For $F=5/2$, the result reads as 
follows (in the basis of states
$\{ 3D_{3/2}^{F=5/2},  3D_{5/2}^{F=5/2} \}$) 
\begin{equation}
\frac{\mathbbm{E}_\tau(F = \tfrac52)}{\mu\mathrm{eV}} =
\left( \begin{array}{cc} 0.000469(5) & -0.000752(7) \\[2ex]
-0.000752(7) & -0.00214(2) \end{array} \right) \,.
\end{equation}
The energy correction for the $3D_{5/2}^{F=7/2}$ state is 
\begin{equation}
E_\tau(F=7/2) = 0.000670(7) \, \mu{\rm eV} \, .
\end{equation}
%
% Conclusions
%
\section{Conclusions}

We have analyzed the contribution of the tensor polarizability of the nucleus
to the energies of the bound states of simple atomic systems, with an emphasis
on muonic deuterium.  The general formalism has been discussed in
Sec.~\ref{sec2}, followed by a detailed analysis of the tensor structure in
Sec.~\ref{sec3}.  The expression for the tensor matrix element that we obtained
is more general than necessary for muonic deuterium, since the spins of the two
constituents $S_1$ and $S_2$ in the expression we have given are arbitrary.
The matrices for the tensor polarizability
energy shifts of muonic deuterium were evaluated in Sec.~\ref{sec4}.
Parametrically, the energy shifts induced by the 
tensor and scalar polarizability
are of the same order of magnitude
[see our general result in Eq.~\eqref{EEmatrix}].
However, one can expect that the 
tensor effect is, nevertheless, numerically
suppressed in a general case, in view
of an (in general) near-spherical 
symmetry of most nuclei, which results in a relatively small 
value of the tensor polarizability constant $\tau_N$. 
For example, for the $2 P_{3/2}^{F=5/2}$ state
of muonic deuterium, the tensor effect is suppressed,
as far as its magnitude is concerned,
in comparison with the scalar effect,
by a factor $1/15$ from the matrix of angular
factors given in Eq.~\eqref{GGmatrix},
and an additional factor of (roughly) $1/20$ in view of the 
suppression of the tensor factor $\tau_d$
in comparison to the scalar factor $\alpha_E$
from Eq.~\eqref{alphaEtaud}. Together with the 
prefactors from Eq.~\eqref{general_polarizability}
($3/2$ for the tensor effect versus $1/2$ for the 
scalar term),
this leads to an effective suppression by a factor 
$\tfrac{1}{20} \times \tfrac{1}{15} \times 3 \approx 1/100$
relative to the scalar effect.

Let us conclude with some remarks regarding 
a possible experimental verification of the 
mixing of $S$ and $D$ states due to the 
tensor polarizability, even if these remarks
are, in part, somewhat speculative. Namely,
in principle, one might envisage to use 
a technique similar to those used for 
the detection of weak-interaction 
effects in atomic systems~\cite{BoBo1974,WoEtAl1997}.
Let us review a few basic ideas.
In weak-interaction experiments,
the parity-violating $P$-state admixture 
to a reference $6S$ state in cesium is measured as 
follows~\cite{BoBo1974,WoEtAl1997}.
One first realizes that, what initially appears to be a ``pure''
$6S$ state really turns out to be superposition of 
$6S$ and an (energetically nearly degenerate) $6P$
state. Taking into account the $P$-state admixture,
it thus becomes possible to drive an 
electric-dipole transition from $6S$ to $7S$, which becomes
allowed (to a certain extent) once the 
(numerically minute, but nonvanishing) parity-violating admixture
to the reference $6S$ state is taken into account.
The amplitude for the $6S \to 7S$ transition 
is a second-order effect involving virtual $nP$ 
states [see Eq.~(3) of Ref.~\cite{BeWi1999}].
If one introduces, deliberately, an additional 
$6P$ admixture to the reference $6S$ state
via a strong, uniform, externally applied  electric field,
then one induces an additional channel for an
electric-dipole transition from the $6S$ to $7S$.
The transition rate from $6S$ to $7S$, which is 
proportional to the square of the coherently 
added amplitudes (weak-interaction$+$laser and
external-field$+$laser) contains
an interference term between the two amplitudes
[see also Eq.~(1) of Ref.~\cite{BeWi1999}
and Eqs.~(1)---(7) of Ref.~\cite{GiWi1986}].
While flipping the sign of the external, electric field,
the interference term flips its sign, which,
for circularly polarized incident light, 
turns out to be proportional to $\vec E_s \cdot \vec k$,
where $\vec E_s$ is the Stark-admixture 
inducing external electric field, 
and $\vec k$ is the laser wave vector.

For the tensor-polarizability-induced 
$D$-state admixture to an $S$ state
in muonic deuterium, one could, in principle,
imagine an analogous procedure.
It would be based on the $3S \to 4F$ transition
in muonic deuterium. 
The analog of the $P$-state admixture
due to the weak-interaction effect is 
is the $D$-state admixture
due to the tensor polarizability of the nucleus.
The analog of the uniform, external,
electric field is a quadrupole field
which, just like the tensor polarizability,
leads to a $D$-state adminxture to the 
reference $S$ state.
Dipole transitions from the $D$-state admixture
to the $4F$ state are allowed.
In full analogy to the weak interaction,
the transition rate from $3S$ to $4F$, which is
proportional to the square of the coherently
added amplitudes (tensor-polarizability$+$laser and
external-quadrupole-field$+$laser) will contain
an interference term between the two amplitudes.
Upon flipping the sign of the quadrupole field,
one should be able to detect the interference term,
and, hence, the tensor polarizability.
A conceivable experimental realization would 
require a sufficient number of muonic deuterium
atoms prepared in the $3S$ state, and the 
experiment would need to proceed within the 
limited radiative lifetime of the $3S$ state
(as well as within the lifetime of the constituent muon).
On the positive side, one observes that the 
tensor polarizability is an effect that 
involves the physics inside the nucleus [on the scale
of a femtometer, see Eq.~\eqref{alphaEtaud}], while the length
scale of weak interactions (mass of the $Z$ boson)
is much smaller and weak-interaction effects
are numerically suppressed in atoms in 
comparison to the nuclear polarizability.
A further, more detailed discussion of 
the experimental challenges and the general
feasibility of such an approach is 
beyond the scope of this paper and is left
as an open problem for future investigations.
The same applies to the evaluation
of the effect discussed here for other electronic
and muonic bound systems with nuclei whose
spin is greater or equal than one, hence,
allowing for the presence of an electric
tensor polarizability.

%
% Acknowledgments
%
\section*{Acknowledgments}

This work was supported by the National Science Foundation through Grants
PHY-2308792 (G.S.A.) and PHY--2513220 (U.D.J.), and by the National Institute
of Standards and Technology Grant 60NANB23D230 (G.S.A.).


\begin{thebibliography}{10}

\bibitem{Pa2011}
K. Pachucki, {\em \relax{Nuclear Structure Corrections in Muonic Deuterium}},
  Phys. Rev. Lett. {\bf 106},  193007  (2011).

\bibitem{PaWi2015}
K. Pachucki and A. Wienczek, {\em \relax{Nuclear Structure in Deuterium}},
  Phys. Rev. A {\bf 91},  040503(R)  (2015).

\bibitem{KaPaYe2018}
M. Kalinowski, K. Pachucki, and V.~A. Yerokhin, {\em \relax{Nuclear-structure
  corrections to the hyperfine splitting in muonic deuterium}},  Phys. Rev. A
  {\bf 98},  062513  (2018).

\bibitem{PaEtAl2024}
K. Pachucki, V. Lensky, F. Hagelstein, S.~S. Li~Muli, S. Bacca, and R. Pohl,
  {\em \relax{Comprehensive theory of the Lamb shift in light muonic atoms}},
  Rev. Mod. Phys. {\bf 96},  015001  (2024).

\bibitem{JiZhPl2024}
C. Ji, X. Zhang, and L. Platter, {\em \relax{Nuclear Structure Effects on
  Hyperfine Splittings in Ordinary and Muonic Deuterium}},  Phys. Rev. Lett.
  {\bf 133},  042502  (2024).

\bibitem{PaLeHa1993}
K. Pachucki, D. Leibfried, and T.~W. H\"{a}nsch, {\em \relax{Nuclear-structure
  correction to the Lamb shift}},  Phys. Rev. A {\bf 48},  R1--R4  (1993).

\bibitem{JeAd2022book}
U.~D. Jentschura and G.~S. Adkins, {\em \relax{Quantum Electrodynamics: Atoms,
  Lasers and Gravity}} (World Scientific, Singapore, 2022).

\bibitem{FrPa2005deuteron}
J.~L. Friar and G.~L. Payne, {\em \relax{Deuteron dipole polarizabilities and
  sum rules}},  Phys. Rev. C {\bf 72},  014004  (2005).

\bibitem{Pi1947iii}
J. Pirenne, {\em \relax{Le champ propre et l'interaction des particules de
  Dirac suivant l'\'{e}lectrodynamique quantique. III. Le syst\`{e}me
  \'{e}lectron-positron}},  Arch. Sci. Phys. Nat. {\bf 29},  265--300  (1947).

\bibitem{Go1929aop}
W. Gordon, {\em \relax{Zur Berechnung der Matrizen beim Wasserstoffatom}},
  Ann. Phys. (Leipzig) {\bf 394},  1031--1056  (1929).

\bibitem{Ma1991appell}
A. Matsumoto, {\em \relax{Multipole Matrix Elements for Hydrogen Atom}},  Phys.
  Scr. {\bf 44},  154--157  (1991).

\bibitem{Ta2002}
V. Tarasov, {\em \relax{W. Gordon's Integral (1929) and its Representations by
  means of Appell's functions $F_2$, $F_1$ and $F_3$}},  Mod. Phys. Lett. B
  {\bf 16},  895--899  (2002).

\bibitem{Wo2024}
S. Wolfram, {\em The Mathematica Book}, 4th Edition (Cambridge University
  Press, Cambridge, UK, 1999); 5th Edition (Wolfram Media Inc., Champaign,
  Illinois, 2003); see also Wolfram Research, Inc., {\em \relax{Mathematica
  14.0}}, (Champaign, Illinois, 2024).

\bibitem{Ed1957}
A.~R. Edmonds, {\em \relax{Angular Momentum in Quantum Mechanics}} (Princeton
  University Press, Princeton, New Jersey, 1957).

\bibitem{PoEtAl2016}
R. Pohl, F. Nez, F.~D. Amaro, F. Biraben, J.~M.~R. Cardoso, D.~S. Covita, A.
  Dax, S. Dhawan, M. Diepold, A. Giesen, A.~L. Gouvea, T. Graf, T.~W.
  H\"{a}nsch, P. Indelicato, L. Julien, P. Knowles, F. Kottmann, E.-O.
  Le~Bigot, Y.-W. Liu, J.~A.~M. Lopes, L. Ludhova, C.~M.~B. Monteiro, F.
  Mulhauser, T. Nebel, P. Rabinowitz, J.~M.~F. dos Santos, L.~A. Schaller, K.
  Schuhmann, C. Schwob, D. Taqqu, J.~F. C.~A. Veloso, and A. Antognini, {\em
  \relax{Laser spectroscopy of muonic deuterium}},  Science {\bf 353},
  669--673  (2016).

\bibitem{LiOhShSo2025}
H. Liu, B. Ohayon, O. Shtaif, and Y. Soreq, {\em \relax{Probing New Hadronic
  Forces with Heavy Exotic Atoms}},  Phys. Rev. Lett. {\bf 135},  131803
  (2025).

\bibitem{KrEtAl2016deuterium}
J.~J. Krauth, M. Diepold, B. Franke, A. Antognini, F. Kottmann, and R. Pohl,
  {\em \relax{Theory of the $n=2$ levels in muonic deuterium}},  Ann. Phys.
  (N.Y.) {\bf 366},  168--196  (2016).

\bibitem{BoBo1974}
M.~A. Bouchiat and C.~C. Bouchiat, {\em \relax{Weak Neutral Currents in Atomic
  Physics}},  Phys. Lett. B {\bf 48},  111--114  (1974).

\bibitem{WoEtAl1997}
C.~S. Wood, S.~C. Bennett, D. Cho, B.~P. Masterson, J.~L. Roberts, C.~E.
  Tanner, and C.~E. Wieman, {\em \relax{Measurement of Parity Nonconservation
  and an Anapole Moment in Cesium}},  Science {\bf 275},  1759--1763  (1997).

\bibitem{BeWi1999}
S.~C. Bennett and C.~E. Wieman, {\em \relax{Measurement of the $6S \to 7S$
  Transition Polarizability in Atomic Cesium and an Improved Test of the
  Standard Model}},  Phys. Rev. Lett. {\bf 82},  2484--2487  (1999), [Errata:
  Phys. Rev. Lett. {\bf 82}, 4153 (1999); Phys. Rev. Lett. {\bf 83}, 889
  (1999)].

\bibitem{GiWi1986}
S.~L. Gilbert and C.~E. Wieman, {\em \relax{Atomic-beam measurement of parity
  nonconservation in cesium}},  Phys. Rev. A {\bf 34},  792--803  (1986).

\end{thebibliography}
\end{document}